# A Fractal Comparison of M.C. Escher's and H. von Koch's Tessellations

B. Van Dusen[1], B.C. Scannell[2] and R.P. Taylor[2]

[1]School of Education, University of Colorado, Boulder, 80309, USA

[2]Physics Department, University of Oregon, Eugene, OR 97403, USA

**Abstract:** M.C. Escher's tessellations have captured the imaginations of both artists and mathematicians. *Circle Limit III* is the most intricate of his tessellations, featuring patterns that repeat at increasingly fine scales. Although his patterns follow a scaling law determined by hyperbolic geometry, his work is often mistakenly described as following fractal geometry. Here, we perform a 'box-counting' scaling analysis on *Circle Limit III* and an equivalent mono-fractal pattern based on a Koch Snowflake. Whereas our analysis highlights the expected visual differences between Escher's hyperbolic patterns and the simple mono-fractal, the analysis also identifies unexpected similarities between Escher's work and the bi-fractal poured paintings of Jackson Pollock.

**1.1 Introduction**

It has become popular to view the spectrum of disciplines as a circle, with mathematics and art lying so far apart that they become neighbors. Two artists are celebrated as proof of this theory - Leonardo da Vinci (1452-1519) and Maurits Escher (1898-1972). Da Vinci combined mathematics and art to search for the possible, resulting in functional designs such as his famous flying machines. In contrast, Escher searched for the impossible. As we will see, he created images by distorting nature's rules.

In this article, we focus on Escher's prints of tessellations [1], of which the woodcut *Circle Limit III* (1958) is the most intricate. Inspired by the Islamic tiles that he saw during a trip to the Alhambra in Spain, Escher took the bold step of incorporating patterns that repeat at many size scales. *Circle Limit III*, shown in Figure 1(a), reflects the mathematical challenge and the troubled artistic road that he took to meet it.

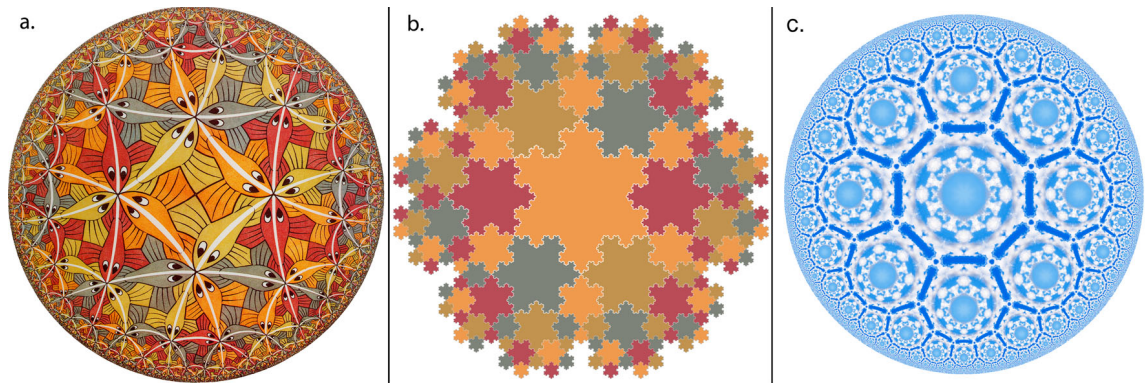

**Figure 1**. a) Escher's *Circle Limit III* (1958). b) A version of a Koch Snowflake generated for direct comparison with Escher's tessellations. c) A hyperbolic tiling generated for comparison with Escher's tessellations.

To achieve the desired visual balance, he insisted that the shrinking patterns converge towards a circular boundary. In Escher's words, the repeating patterns emerge from the circular boundary "like rockets", flowing along curved trajectories until they "lose themselves" once again at the boundary [2]. Making his patterns fit together required considerable thought and a helping hand from mathematics. After several flawed attempts, Escher finally found the solution in an article written several years earlier by the British geometer H.S.M. Coxeter [3, 4].

These flowing patterns have captured the imaginations of both artists and mathematicians for over half a century. Yet along the way, their connection with nature has fallen by the way side. His work is often presented as an elegant solution to a purely academic exercise of mathematics - a clever visual game. In fact, Escher's interest lay in the fundamental properties of patterns that appear in the real world. He declared: "We are not playing a game of imaginings – we are conscious of living in a material three dimensional reality." [2].

Escher's artistic interest in the physical world is emphasized by the sketches of trees that he completed in the same era as *Circle Limit III* [1]. These sketches demonstrate how branch patterns repeat at different size scales and how they become distorted when reflected in the rippled surface of a pond. Given his quest for distortion, it would be surprising to find that *Circle Limit III* was an exact replication of nature's patterns. In this article, we focus on the fact that his patterns shrink at a different rate than those found in nature. The artist used the hyperbolic geometry described in Coxeter's article [3, 4], rather than the fractal geometry that describes natural patterns [5].

To highlight the visual differences between Escher's hyperbolic geometry and fractal geometry, we will apply a computer scaling analysis to *Circle Limit III* and an equally famous pattern from fractal geometry – the Koch Snowflake. The Koch curve was created fifty years earlier by the mathematician Niels Fabian Helge von Koch (1870-1924). Koch's curve consisted of triangles that repeat at increasingly fine scales, so building up the edge of a snowflake [6].

To facilitate a direct comparison with Escher's patterns, we created the snowflake pattern shown in Figure 1(b) and the hyperbolic tiling of clouds shown in Figure 1(c). Whereas the snowflake pattern features interlocking tessellations similar to *Circle Limit III*, the scaling properties of these tessellations are set by the triangles of the central snowflake, which are fractal rather than hyperbolic. The cloud pattern was created using commercially available software and shows an exact hyperbolic pattern. By examining the results of the scaling analysis, we discuss the visual implications of Escher's deviations from the fractal and hyperbolic scaling of nature's shapes.

We note that significant formal mathematical comparisons between fractal and hyperbolic geometries have been made previously, notably by Stratmann [7]. Here we use a different analytical technique, by comparing the edge patterns of the three art works of Figure 1 employing a traditional fractal analysis called the "box counting" technique. This technique is particularly appropriate because visual perception studies highlight the importance of edges for distinguishing patterns [8]. Furthermore, the box-counting method assesses the space-filling characteristics of these edges at different size scales. Space-filling properties have been found to be an important aesthetic quality for other forms of abstract art featuring multi-scale patterns, most notably the poured paintings of

American artist Jackson Pollock (1912-1956) [9]. Whereas our analysis highlights the expected visual differences between Escher's hyperbolic patterns and the simple Koch fractal, the analysis also identifies unexpected visual similarities between Escher's work and the fractal poured paintings of Pollock.

**1.2 Escher's Construction of Multi-scale Tessellations**

Tessellations are patterns that fill the surface of a plane without any overlaps or gaps. Most tessellation designs in art feature component tiles which are identical in size. In the four piece series *Circle Limit*, Escher instead created tessellations that appear to be 'going to infinity' as they approach the edge of a circle: in other words, their size diminishes towards the infinitesimally small as the edge is approached. To create this visual effect, he had to generate tile shapes that are not typically tiled together.

When taking the basic shapes that Escher used in his *Circle Limit* pieces and tiling them together they don't create a flat surface. Instead of fitting together evenly in two dimensions, the pieces form the hyperbolic geometry shown in Figure 2. A hyperbolic geometry looks much like a saddle: on one axis, the surface rises upward from the origin (symbolized by the dot) and on the other axis the surface drops downward. When this hyperbolic surface is viewed directly from above, it appears to spread out indefinitely and any tiles 'drawn' on the surface become more distorted the further they get from the origin. Thus, although all the tiles have equal size on the hyperbolic surface, they appear to shrink towards the edge when viewed this way.

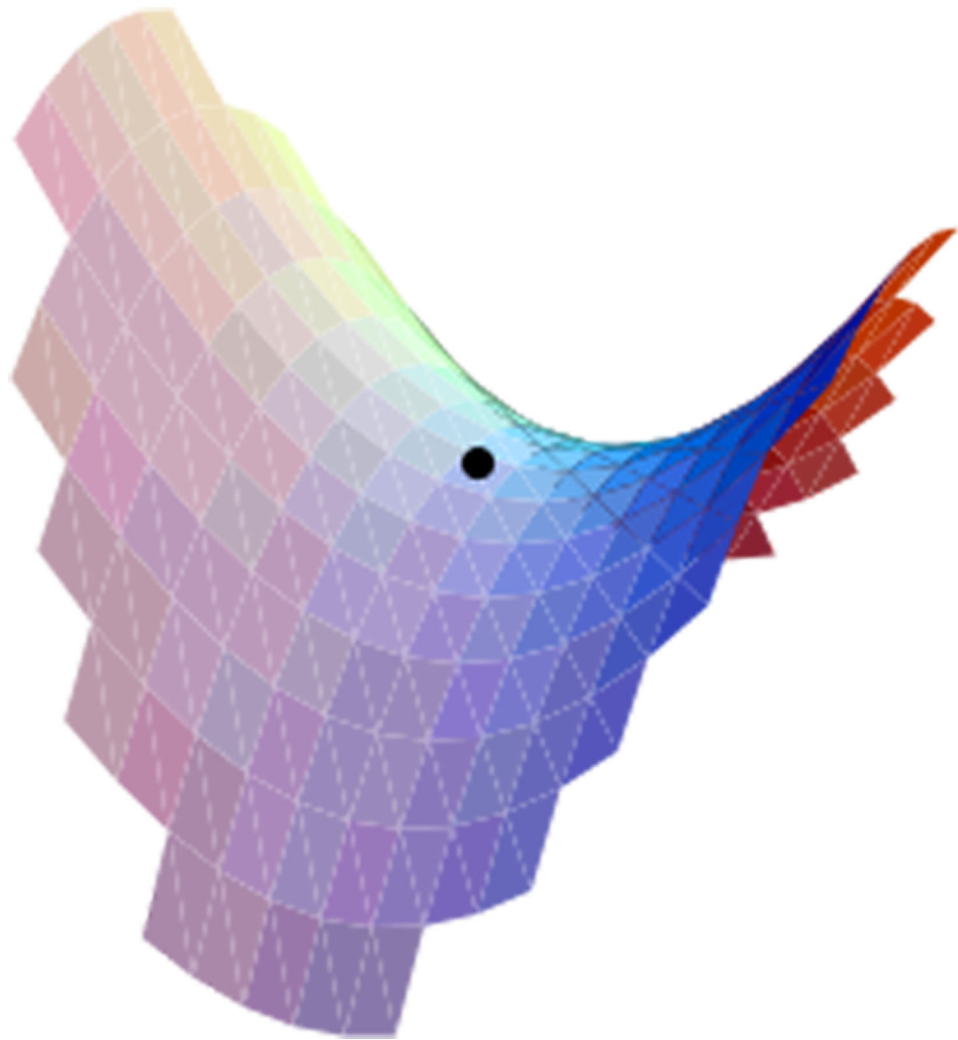

**Figure 2.** Hyperbolic geometries form the shape of a saddle with the surface rising along one axis and dropping along the other axis.

To make the entire surface viewable, Escher translated the surface onto a Poincaré disk [10]. A Poincaré disk is a circle that represents an infinite region of space. As the circular edge is approached, the images diminish at such a rate that they appear to be getting both infinitely smaller and closer to the circle's edge, without ever actually reaching it. By using this Poincaré disk model, Escher was able to give the impression of an infinite array of tile images within a limited space, and unlike other disk models; the shape of the tiles stays recognizable as they approach the circular boundary.

In his first attempt at using the Poincaré disk model, Escher was dissatisfied with his final product, *Circle Limit I* (1958)**.** He felt that it lacked "traffic flow" and unity of color in each row [1]. Escher was much happier with what is referred to as his most accomplished *Circle Limit* piece, *Circle Limit III*:

> "*Circle Limit I*, being a first attempt, displays all sorts of shortcomings... There is no continuity, no "traffic flow," nor unity of colour in each row... In the coloured woodcut *Circle Limit III*, the shortcomings of *Circle Limit I* are largely eliminated. We now have none but "through traffic" series, and all the fish belonging to one series have the same colour and swim after each other head to tail along a circular route from edge to edge... Four colours are needed so that each row can be in complete contrast to its surroundings." [1]

In *Circle Limit III*, Escher tessellates four different colored fish. Each fish has one of its fins touching the fins of three fish and its other fin touching the fins of two fish. Each fish nose touches two other fishes' noses and three tails. As shown in Figures 3(a) and 3(b), this pattern is a tessellation of octagons. Each octagon is met at its corner with two other octagons.

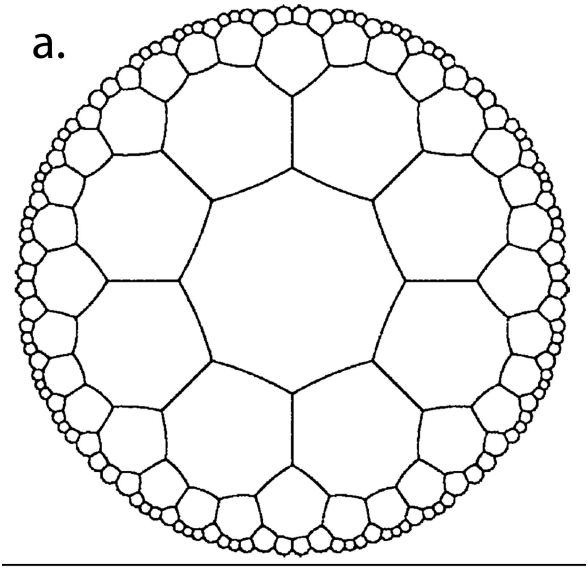
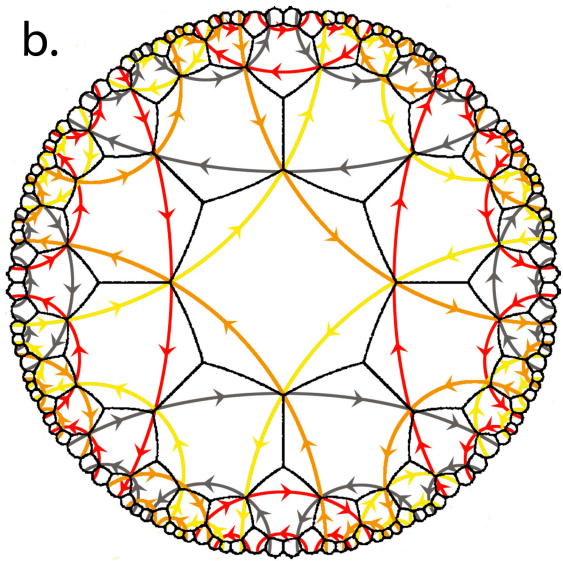
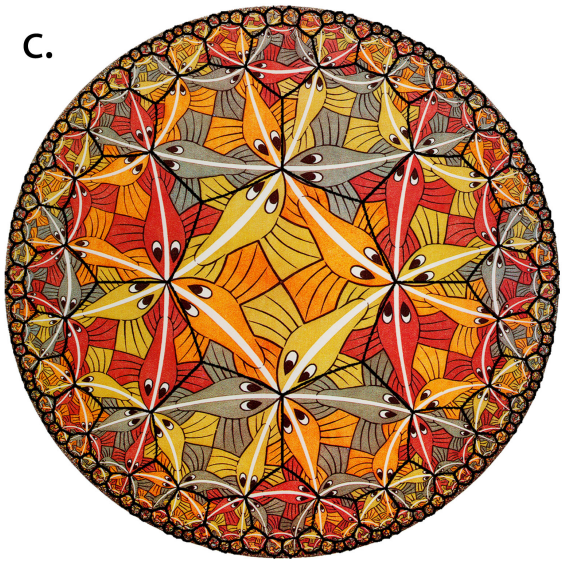

**Figure 3**. a) Escher's octagonal pattern used in *Circle Limit III*. b) The octagonal pattern overlaying the circular arcs created by the fishes' spines, color-coded to each type of fish. c) The octagonal pattern laid over *Circle Limit III*.

When a hyperbolic tessellation is mapped onto a Poincaré disk, straight lines become circular arcs that are orthogonal to the bounding circle. This is demonstrated with the 'flow lines' shown in Figure 3(b). In a mistake that appears to be unbeknownst to Escher himself, unlike his other Circle Limit pieces, *Circle Limit III*'s arcs aren't true hyperbolic arcs. When analyzed by Coxeter, the white circular arcs approach the boundary of the disk at 80°, not the 90° of hyperbolic line [4]. Despite this minor aberration in the pattern, it still gives a viewer the strong sense of the pattern approaching the infinite. Further details of Escher's hyperbolic design can be found elsewhere [11].

**1.3 Fractal Analysis of *Circle Limit III***

The tessellations within Escher's *Circle Limit* series are often mistaken for fractal images. Like fractals, the *Circle Limit* series have self-similar patterns that recur at increasingly finer scales. However, Escher's tessellations decrease at a hyperbolic-like rate, rather than the power law rate at which fractals decrease [5, 9, 12]. The visual difference of these three scaling rates is highlighted in Figure 4, where the edge patterns for the three tessellations designs have been isolated.

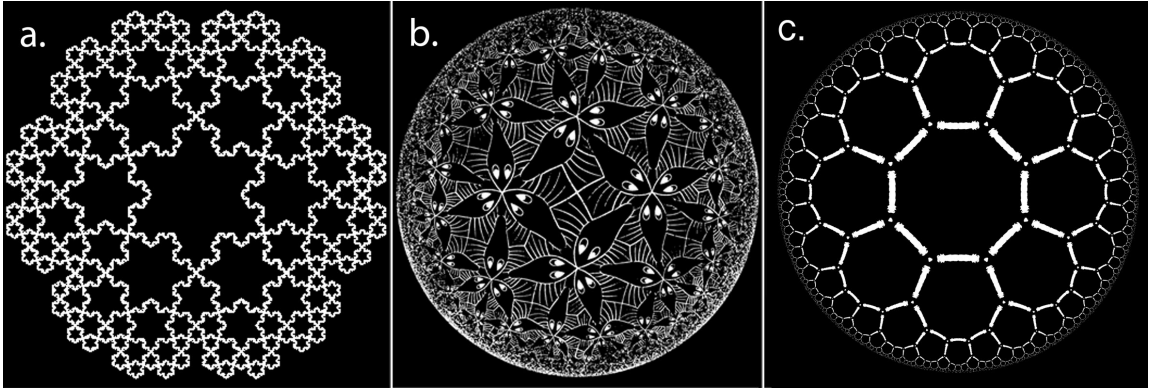

**Figure 4.** The edge patterns of a) the Koch Snowflake, b) *Circle Limit III*, and c) a true hyperbolic tiling.

This isolation of the edge patterns also allows their scaling behavior to be analyzed using the "box counting" technique [9, 12]. Adopting this technique, the image of white edges is covered with a computer-generated grid of identical squares (or "boxes"), as shown in Figure 5. By analyzing which of the squares are "occupied" (i.e., contains a part of the white edge pattern) and which are "empty" (shaded blue in Figure 5), the statistical qualities of the edge pattern can be calculated.

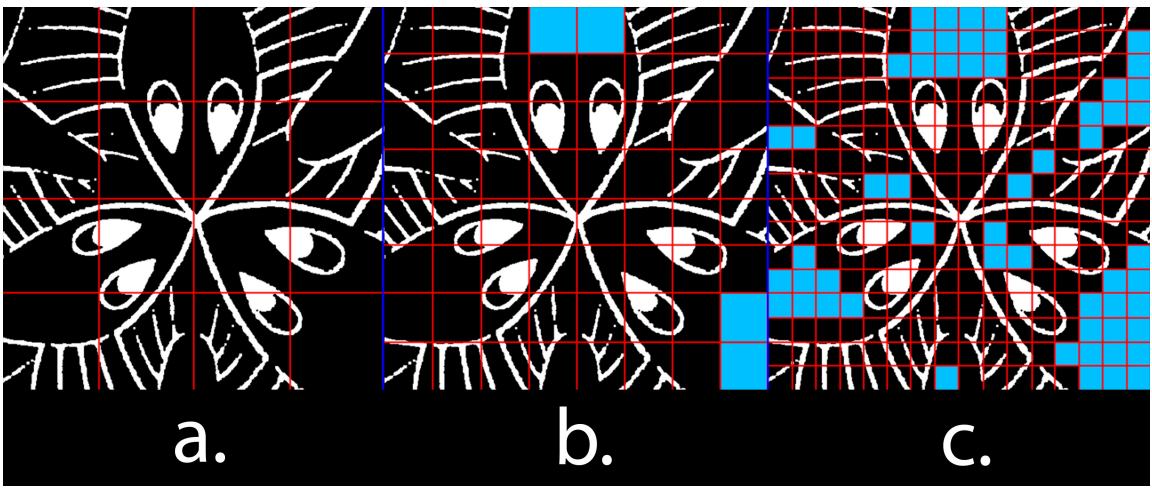

**Figure 5.** A section of *Circle Limit III* overlaid with a grid of boxes. Boxes containing white pixels are counted in the box counting procedure. This count is repeated for increasingly fine boxes sizes (a, b, c).

Reducing the square size in the grid is equivalent to looking at the pattern at a finer magnification. Thus, in this way, the pattern's statistical qualities can be compared at different magnifications. Specifically, if $N$, the number of occupied squares, is counted as a function of $L$, the width of each square, then for fractal behavior $N(L)$ scales according to the power law relationship $N(L) \sim L^{-D}$ [5, 9, 12]. The exponent $D$ is called the fractal dimension and its value can be extracted from the gradient of the "scaling plot" of $\log(N)$ plotted against $\log(1/L)$.

The scaling plots generated by the box-counting technique are shown in Figure 6. The analysis is performed over a magnification range lying between coarse and fine scale cut-offs. The coarse scale cut-off is set by the box size ($L \sim 6$cm) at which the grid has less than 50 boxes: at larger $L$ values, there are insufficient boxes to ensure reliable counting statistics [9]. The fine scale cut-off is set by the box size ($L \sim 0.5$mm) corresponding to approximately 3 pixels of the image: at smaller box sizes, the count is compromised by the resolution limit of the image. Between these two cut-offs, the scaling plot for the Koch Snowflake displays the straight power-law line expected for a fractal pattern. In contrast, the data for *Circle Limit III* fails to condense onto a straight line. To interpret the visual significance of this curved line, we need to consider the importance of the fractal's power law line in more detail.

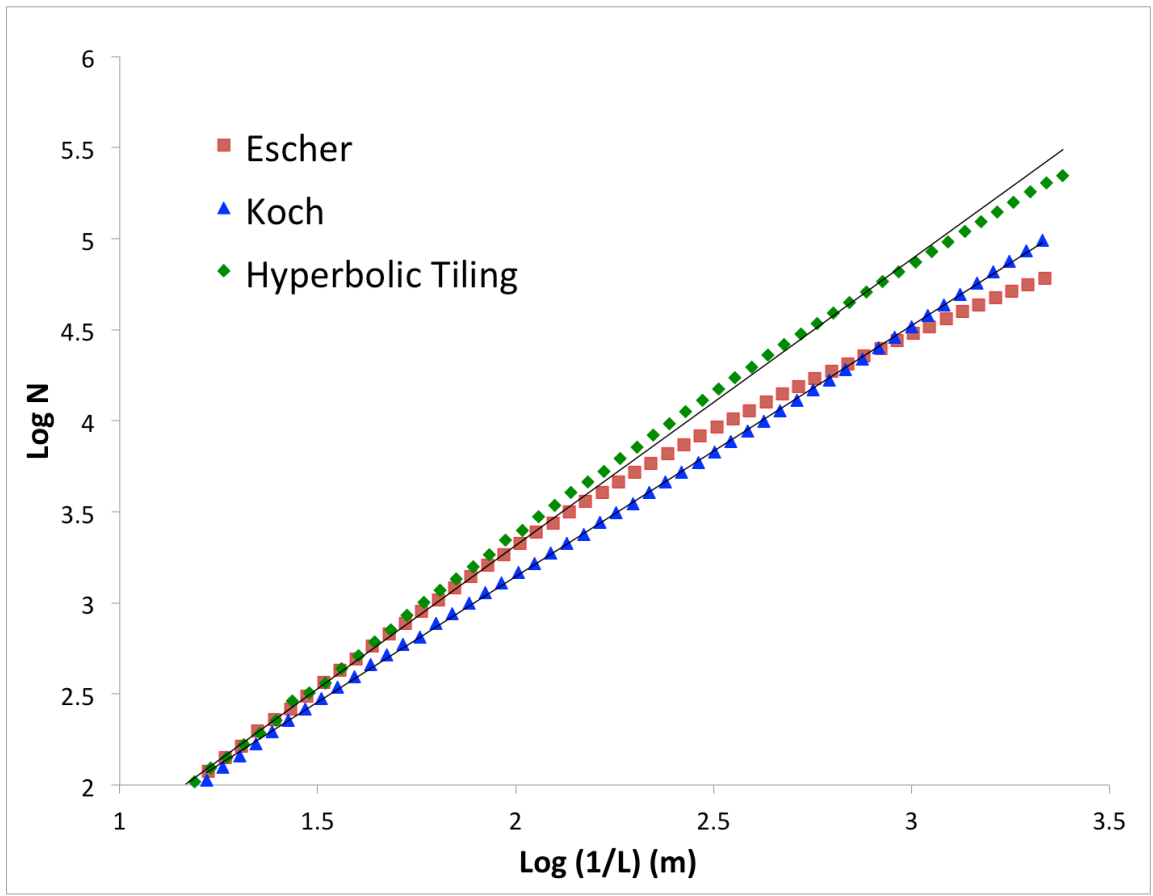

**Figure 6.** Scaling plots obtained from the box counting analysis of the Koch Snowflake (blue data), *Circle Limit III* (red data), and a true hyperbolic tiling (green data). The analyzed images had dimensions 42 cm by 42 cm and resolution of 63 pixels per cm.

The power law line generates the scale-invariant properties that are central to fractal geometry. It also quantifies the crucial role played by *D* in determining the pattern's visual appearance. *D* corresponds to the gradient of the scaling plot. Therefore, a high *D* value is a signature of a large *N* value at small *L* and reflects the fact that many small boxes are being filled by fine structure. This can be seen, for example, for the set of Koch curves shown in Figure 7. The fine features play a more dominant 'space coverage' role for the high *D* pattern than for the low *D* pattern. For fractals described by a low *D* value, the patterns observed at different magnifications repeat in a way that builds a relatively smooth-looking shape compared to the complex, detailed structure of high $D$ patterns.

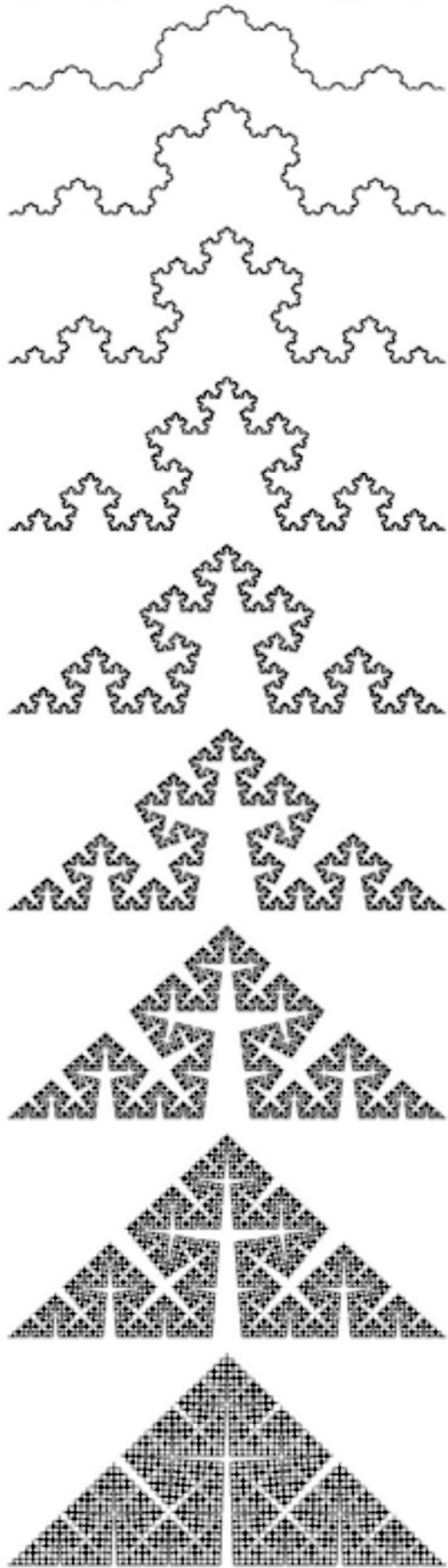

**Figure 7.** Koch curves with *D* values ranging from 1.1 (top) to 1.9 (bottom).

Perception experiments confirm that raising the *D* value of the fractal pattern increases its perceived roughness and complexity [13, 14, 15]. Clearly, *D* is a highly appropriate tool for quantifying fractal complexity. Traditional measures of visual patterns quantify complexity in terms of the ratio of fine structure to course structure. *D* goes further by quantifying the relative contributions of the fractal structure at all the intermediate magnifications between the course and fine scales.

As expected from the previous formal comparisons of hyperbolic and fractal geometry [7], Escher's tessellations deviate from the straight data line of fractal geometry. The importance of Figure 6 lies in the fact that it highlights the precise deviation in terms of the space-coverage of the pattern at different scales. How, then, do we interpret the visual impact quantified by the curved data line for *Circle Limit III*? Because of its curvature, the data line can't be quantified by a *D* value. Nevertheless, the steepness of the line holds the same visual consequences as for a fractal pattern: steeper gradients indicate that the space-coverage of the pattern is changing at a faster rate when zooming into finer size scales. Intriguingly, the rate characterizing *Circle Limit III* can be seen to 'weave' around the constant rate set by the fractal geometry of the Koch pattern. More precisely, the rate is steeper than the Koch fractal at coarse scales but then shallower at the finer scales. A visual inspection of Figures 1 and 4 confirms that this is indeed the case.

As noted previously, Figure 6 shows that Escher's work doesn't create a constant line that is attributed to fractals. However, it is also worth noting that Escher's curve deviates from a true hyperbolic geometry. Our box counting program is able to show that true hyperbolic geometries have a slight curve in their data, while Escher's data has a more significant deviation from a straight line.

**1.4 Conclusions**

In contrast to the traditional view of Escher's achievements in terms of abstract mathematics, in this article we have instead concentrated on his interest in nature's patterns. To do this, we presented a comparison of Escher's repeating tessellations and those of fractal geometry, and so showed that his patterns deviated from nature's rules of scaling as well as those of hyperbolic geometry.

Intriguingly, *Circle Limit III* was created many years before B.B. Mandelbrot's "Fractal Geometry of Nature" made nature's scaling properties well known [5]. Therefore, the intriguing question raised in this article is the extent to which Escher knew about the distortions of nature that he captured so precisely in much of his art. Our results indicate that Escher knew of the mild distortion from fractals that hyperbolic tiling produced, but that he chose to perform a modified hyperbolic tiling to produce an even larger distortion.

A common theme that emerges from fractal studies is that artists were mimicking nature's fractals prior to their scientific discovery. Perhaps this is a consequence of the intimacy resulting from art and mathematics being so far away and yet so close. Escher's own words hint at knowledge of his distortions of nature: "The reality around us… is too common, too dull, too ordinary for us. We hanker after the unnatural or supernatural, that which does not exist, a miracle" [1]. Perhaps he achieved this miracle in what he referred to as the "deep, deep infinity" of his repeating patterns.

What, though, are the aesthetic implications of Escher's deviations from fractal geometry? Does the fact that *Circle Limit III* diminishes at a varying rate similar to the curvature of the hyperbolic surface hold important potential for visual perception studies? Previous perception experiments have concentrated on patterns generated by the constant scaling rate of simple fractals [16, 17, 18] and so a detailed quantification of the aesthetic impact of the curved rate will have to await future experiments. Nevertheless, an initial observation can be made by comparing *Circle Limit III* to the poured paintings of Jackson Pollock. Although Pollock's paintings are fractal, and are therefore described by a fundamentally different geometry to Escher's hyperbolic patterns, the two art works have an unexpected shared scaling characteristic, as follows. Pollock's paintings have been shown to be 'bi-fractal' – patterns at the large size scales created by his body motions are quantified by a high $D$ value, while the splatter patterns observed at finer scales have a low $D$ value [9]. The visual consequence of this bi-fractal behavior is similar to that of the hyperbolic curve of Escher's work – a steep rate of space-coverage at large scales followed by a shallow rate at finer scales. Although we emphasize the preliminary character of this observation, it is nevertheless intriguing to hypothesize that both artists regarded the constant scaling rate of 'mono' (single $D$) fractals, such as the Koch snowflake, to be too monotonous to exhibit aesthetic appeal.